\documentclass{article}

\usepackage[final, nonatbib]{neurips_2020_ml4ps}

\usepackage[utf8]{inputenc} 
\usepackage[T1]{fontenc}    
\usepackage{hyperref}       
\usepackage{url}            
\usepackage{booktabs}       
\usepackage{amsfonts}       
\usepackage{nicefrac}       
\usepackage{microtype}      
\usepackage{mathtools}
\usepackage{graphicx}
\usepackage{subcaption}
\usepackage{svg}
\usepackage{caption}
\usepackage{epstopdf}
\usepackage{color}
\usepackage{amsmath}
\usepackage{xcolor}
\usepackage{svg}

\title{AGNet: Weighing Black Holes with Machine Learning}

\author{
  Joshua Yao-Yu Lin\thanks{equal contribution} \\
  University of Illinois at Urbana-Champaign\\
  NCSA \\
  Urbana, IL, USA \\
  \texttt{yaoyuyl2@illinois.edu} \\
  \And
  Sneh Pandya$^*$ \\
  University of Illinois at Urbana-Champaign\\
  NCSA \\
  Urbana, IL, USA \\
  \texttt{snehjp2@illinois.edu} \\
  \And
  Devanshi Pratap \\
  Microsoft\\
  University of Illinois at Urbana-Champaign\\
  NCSA \\
  Urbana, IL, USA \\
  \texttt{pratapdevanshi@gmail.com} \\
   \And
  Xin Liu \\
  University of Illinois at Urbana-Champaign\\
  NCSA \\
  Urbana, IL, USA \\
  \texttt{xinliuxl@illinois.edu} \\
   \And
  Matias Carrasco Kind \\
  National Center for Supercomputing Applications (NCSA)\\
  University of Illinois at Urbana-Champaign\\
  Urbana, IL, USA \\
  \texttt{mcarras2@illinois.edu} \\
}

\begin{document}

\maketitle

\begin{abstract}
Supermassive black holes (SMBHs) are ubiquitously found at the centers of most galaxies. Measuring SMBH mass is important for understanding the origin and evolution of SMBHs. However, traditional methods require spectral data which is expensive to gather. To solve this problem, we present an algorithm that weighs SMBHs using quasar light time series, circumventing the need for expensive spectra. We train, validate, and test neural networks that directly learn from the Sloan Digital Sky Survey (SDSS) Stripe 82 data for a sample of $9,038$ spectroscopically confirmed quasars to map out the nonlinear encoding between black hole mass and multi-color optical light curves. We find a 1$\sigma$ scatter of 0.35 dex between the predicted mass and the fiducial virial mass based on SDSS single-epoch spectra. Our results have direct implications for efficient applications with future observations from the Vera Rubin Observatory.
\end{abstract}

\section{Introduction}

Supermassive black holes (SMBHs) with masses of millions to tens of billions times the mass of the Sun are commonly found at the hearts of massive galaxies \cite{KormendyHo2013}. While black holes themselves are invisible, as light cannot escape them, the associated phenomena are visible. These actively feeding SMBH are known as Active Galactic Nuclei (AGN). The most dramatic of these is called a quasar. Quasars are among the most powerful and distant objects in the universe \cite{Schmidt1968}. The glow of matter as it falls into SMBHs is what makes quasars so bright \cite{Rees1984}. 

Quasars provide a window to study how a SMBH grows with time \cite{Soltan1982}. Because of their brightness, it is possible to detect quasars almost close to the edge of the observable universe \cite{Banados2018}. They offer a ``standard candle'' to study the expansion history of the universe to understand the nature of Dark Energy \cite{King2014,Dultzin2020} -- arguably the biggest mystery in contemporary astrophysics. 

Measuring SMBH mass and redshift is important for understanding the origin and evolution of quasars. However, traditional methods require spectral data which is highly expensive; the existing $\sim$0.5 million mass estimates represent about 30 years' worth of community efforts \cite{Shen2013,rakshit2020spectral}. The Vera C. Rubin Observatory \cite{Ivezic2019} Legacy Survey of Space and time (LSST) will discover $\sim$17 million quasars\footnote{https://www.lsst.org/sites/default/files/docs/sciencebook/SB\_10.pdf}. A much more efficient approach for estimating SMBH mass is needed to maximize LSST AGN science, which is, however, still lacking.


Here we present a new approach to solve the problem based on Machine Learning (ML). ML has been applied for many applications in astronomy \cite{Cabrera-Vives2017,Charnock2017,Kim2017,George2018a,Huang2018,Lanusse2018,Ribli2018,Burke2019, yao2020feature}. Recently, ML has been employed to classify quasars and predict their cosmological redshift \cite{Pasquet-Itam2018} using data from the Sloan Digital Sky Survey \cite{york2000sloan} 
Autoencoders have also been used to extract time series features to model the quasar variability and estimate quasar mass \cite{tachibana2020deep}.

The current work represents the first ML application for predicting quasar mass with multi-band time series. Our ML approach is well motivated. There is empirical evidence and theoretical reasons to believe that the quasar light curve encodes physical information about its SMBH mass \cite{kelly2009variations, macleod2010modeling}. However, the encoding is nonlinear and difficult to model using standard statistics. We train neural networks that directly learn from the data to map out the nonlinear encoding. Our approach is fundamentally different from previous methods. The networks directly weigh SMBHs using quasar light curves, which are much cheaper to collect for a large sample. Our scheme is directly applicable for the Vera C. Rubin Observatory LSST \cite{Ivezic2019}, achieving a high efficiency by circumventing the need for expensive spectroscopic observations.


\section{Data and Method}\label{sec:method}

\subsection{Data}

\subsubsection{SDSS Stripe 82 Light Curves}\label{subsubsec:s82}

We adopt multi-color light curves from the SDSS Stripe 82 
as our training and testing data. Our sample consists of $\sim$10,000 quasars. 
We take the mean of the \emph{ugriz} magnitudes 
as features. 
We adopt the additional time series features from the Damped-Random-Walk model fitting parameters \cite{macleod2010modeling}, such as the variability timescale ($\tau$) and variance, ($\sigma$). 
Time series features have shown to be useful in predicting quasar redshift \cite{Pasquet-Itam2018} using the Python library Feature Analysis Time Series (FATS) \cite{nun2015fats}.


\subsubsection{Virial Black Hole Mass and Spectroscopic Redshift}\label{subsubsec:virialmass}

We assume the virial SMBH mass estimates from 
\cite{Shen2011} and the spectroscopic redshifts as the ground truth. 
Cosmological redshift is the distortion of light caused by the expansion of space. 

\subsection{Data Preprocessing}

We first clean the sample by removing quasars with no reliable virial mass estimates (e.g. due to low spectroscopic data quality).  
In addition to using the mean \emph{ugriz} bands as features, we compute the colors ($u-g$, $g-r$, $r-i$, $i-z$, $z-u$) as features for a quasar. 
The colors are more robust features than the individual bands in that they provide more information regarding the quasar's 
spectral energy distribution (SED) and temperature. We further use cosmological redshift and K-corrected $i$-band magnitude ($M_i$) as features in predicting mass to provide enough information for the network to infer the intrinsic luminosity of the quasar.  
To standardize the effect of our features in training we apply the scikit-learn \emph{StandardScaler} which removes the mean and scales the data to unit variance.  We split our baseline dataset into an 85$\%$ training and 15$\%$ testing set.


\begin{figure}
 \centering
  \includegraphics[width = \textwidth]{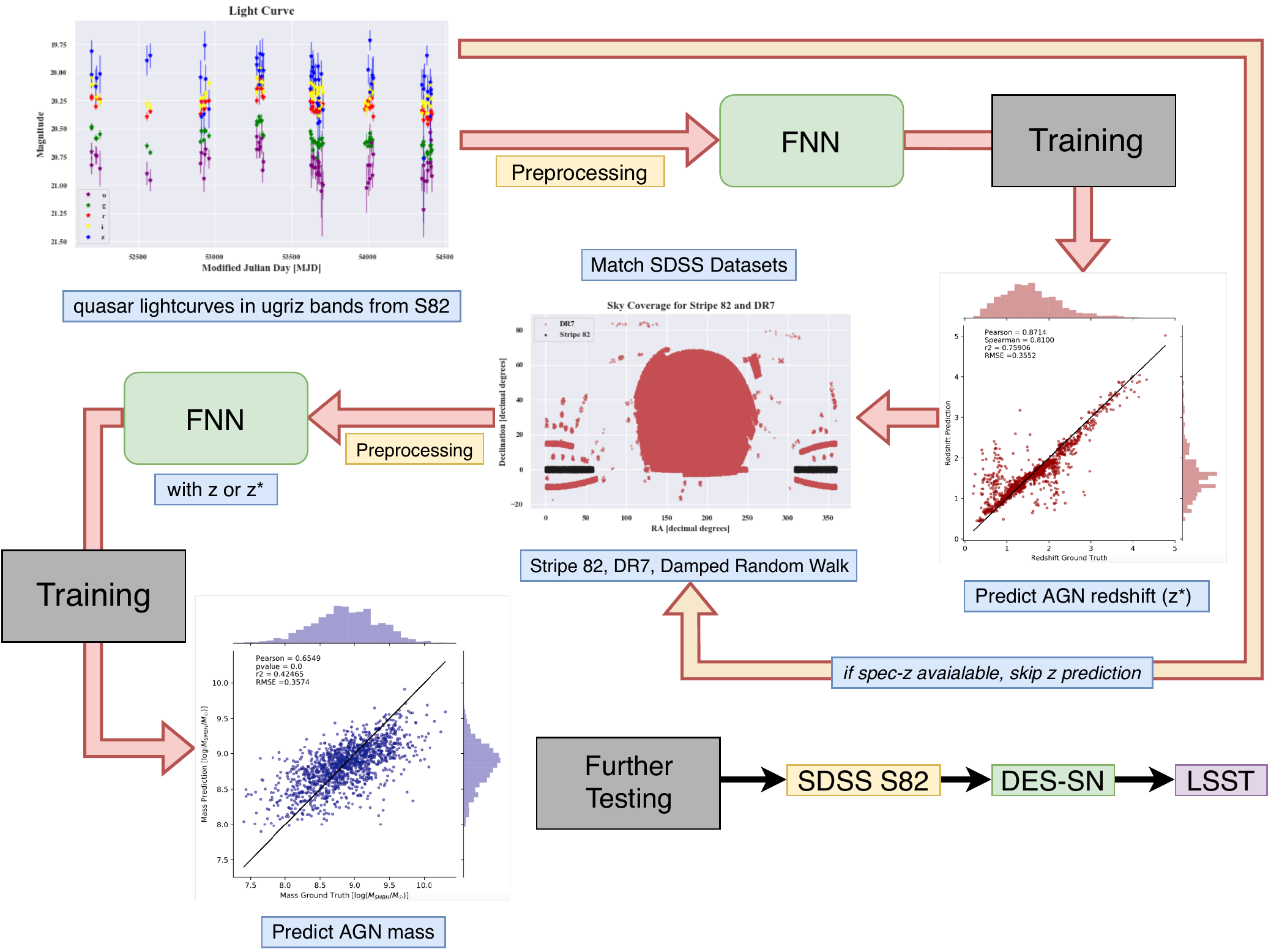}
    \caption{AGNet Implementation Flowchart.}
  \label{fig:flowchart}
\end{figure}

\subsection{Neural Networks}

We use Feedforward Neural Networks (FNN) to predict SMBH mass and redshift. Our FNN architecture for SMBH mass prediction features a 9-neuron input layer, followed by 5 hidden layers with 64 neurons each. Our architecture for redshift is identical with exception of a 10-neuron input layer. For the mass implementation we use the SDSS colors, redshift, $M_i$, and $\tau$, $\sigma$ as features, and just using the SDSS magnitudes and colors as features in predicting redshift.

A diagram of our network architecture for mass is shown in Figure 3, with each node representing four neurons. We investigated multiple different combinations of base-8 neuron architectures, and concluded that 64 neurons gave the best results. We followed a similar process to determine that network performance did not increase greatly for hidden layers greater than five. We use ReLU activation on all neurons \cite{agarap2018relu} after each fully connected layer. 

We train our neural network with gradient descent based AdamW optimizer, an adaptive learning rate optimization algorithm \cite{kingma2014adam, loshchilov2017decoupled}, with default learning rate set as $0.01$.  We optimize our FNN with the SmoothL1 loss function given as:
\begin{equation}\label{eq:loss}
\mathcal{L} = 
\begin{cases}
                $0.5$(x-y){^2} , & \lvert x-y \rvert $ < $ $1$  \\
                \lvert x-y \rvert - $0.5$ , & \text{otherwise}
                
\end{cases}
\end{equation}
where $x$ is the network prediction and $y$ is the ground truth value.  We train for 50 epochs.

Our network currently operates with $\sim$20,000 parameters. We investigated a different neural network architecture with less parameters to reduce the chance of overfitting associated with the number of network parameters.

We considered other loss functions, notably mean squared error, but decided on SmoothL1 because of better performance. We previously investigated RNN and CNN architectures using image and tensor transformations to transform light curve image data into 2D numpy tensors. We incorporated transfer learning and applied ResNet18 and Google EfficientNet architectures \cite{he2016deep, tan2019efficientnet}, however due to limitations in data sample size and irregular gaps in the light curve data, we found that data preprocessing and engineering helped the neural networks to perform better. We additionally tested a variety of loss functions, feature scalers, and other hyperparameters for tuning.

\begin{figure}
 \centering
  \includegraphics[width=.6\textwidth]{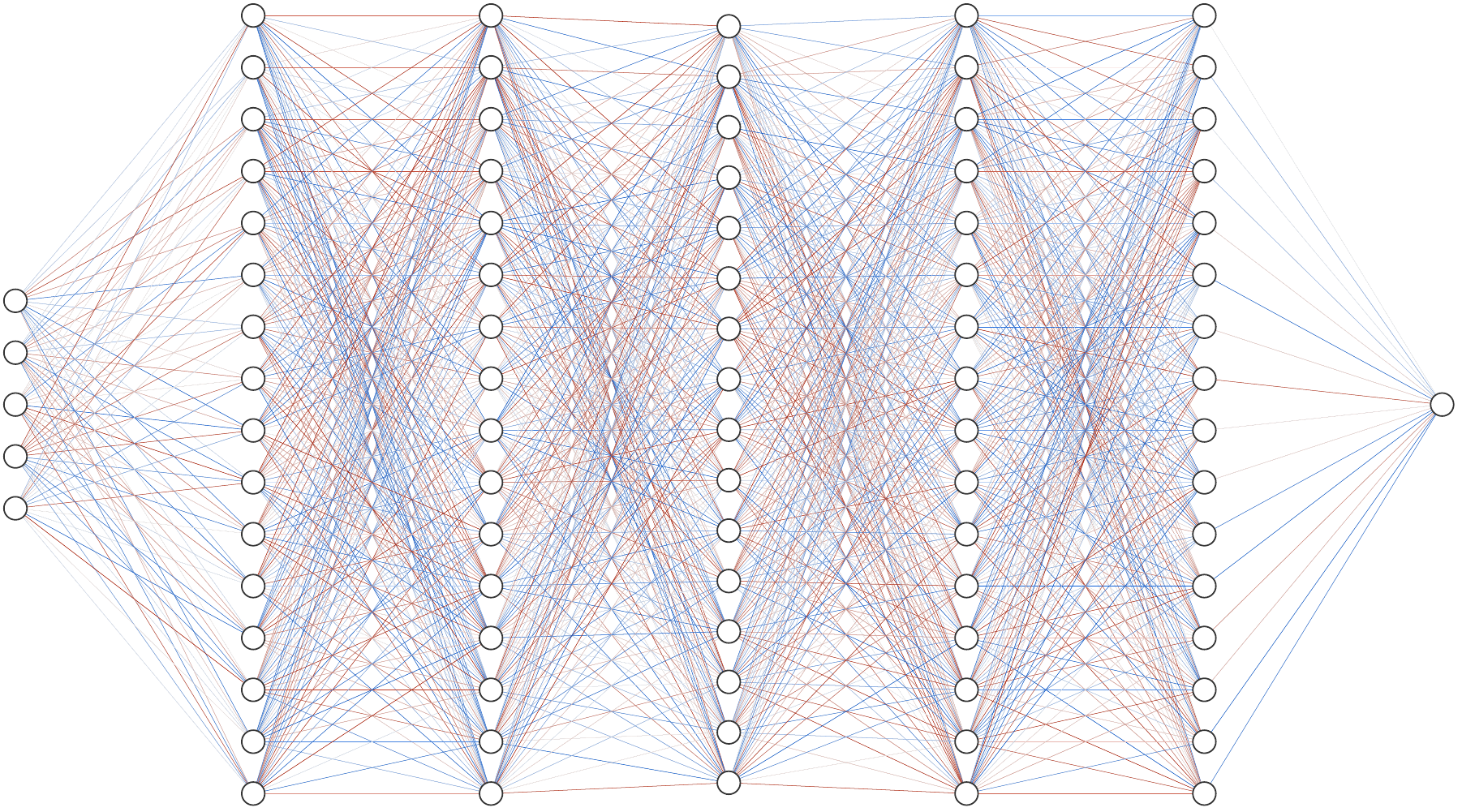}
    \caption{Network Architecture (made using NN-SVG).  Each hidden layer neuron corresponds to four neurons. See Table \ref{tab:summary_stats} for dimension of input layer.  Output neuron can be AGN redshift or AGN SMBH mass depending on implementation.}
  \label{fig:pipeline}
\end{figure} 

\begin{figure}[!tbp]
  \centering
  \subfloat{\includegraphics[width=0.475\textwidth]{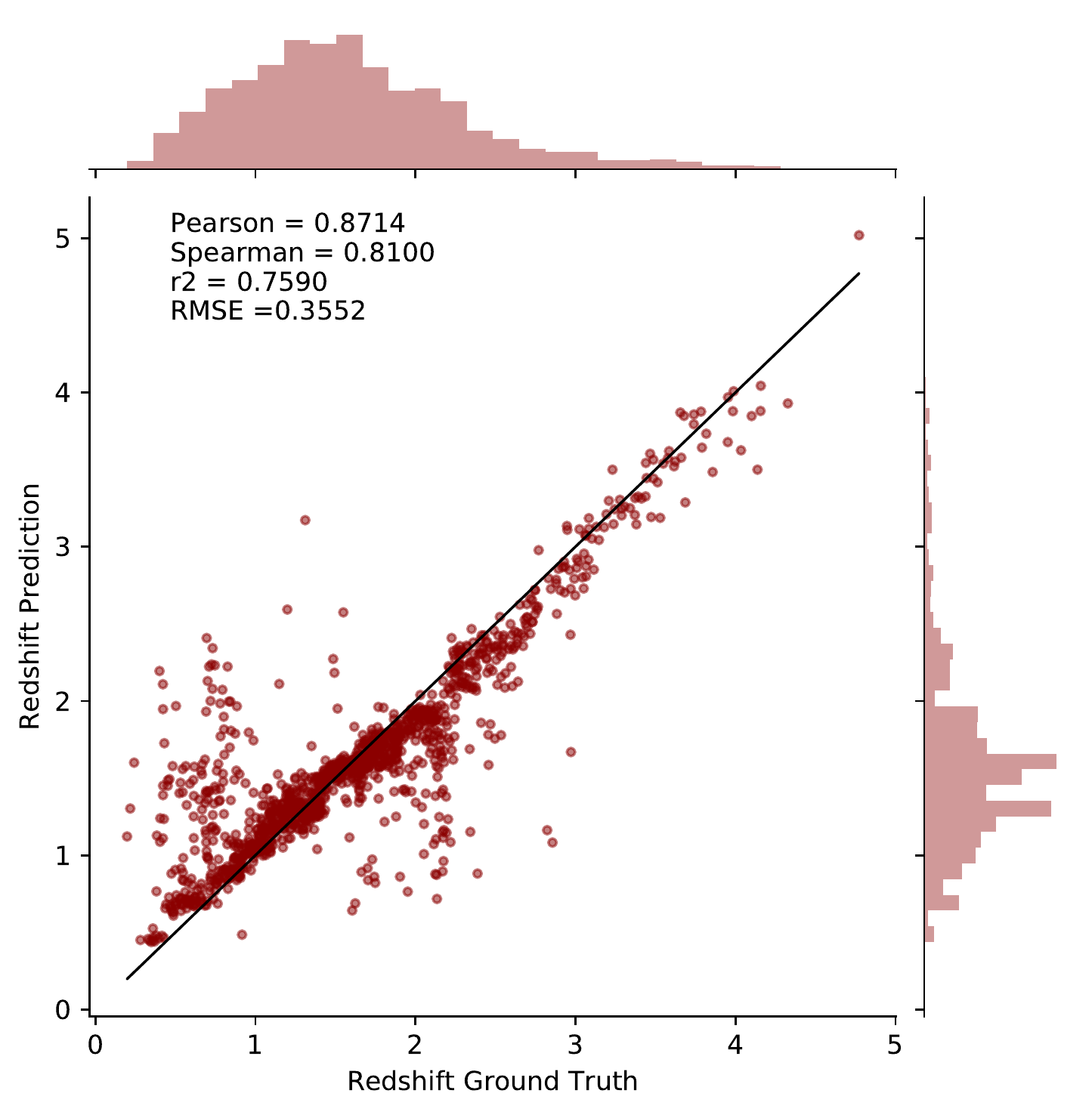}\label{fig:f1}}
  \hfill
  \subfloat{\includegraphics[width=0.475\textwidth]{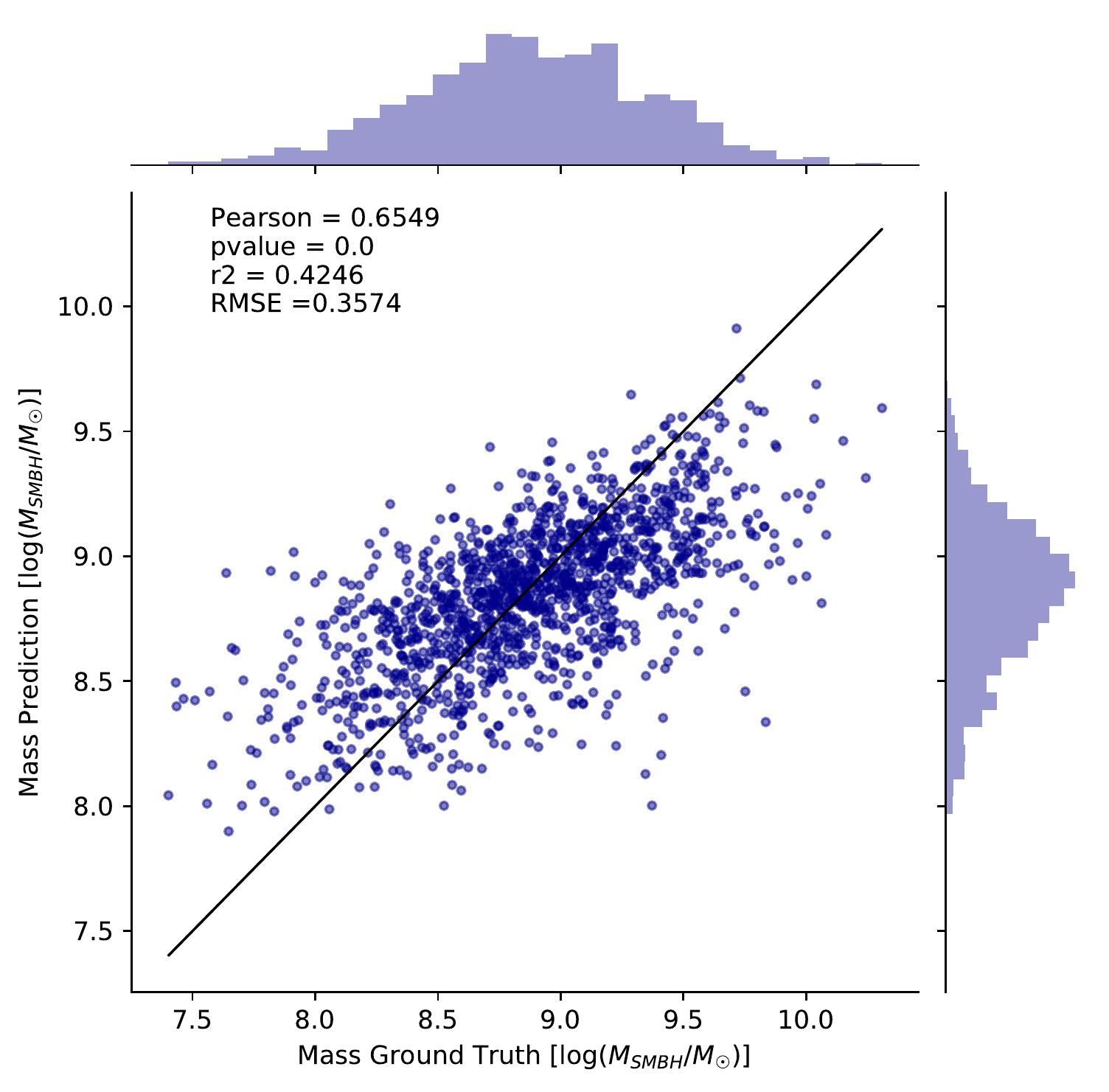}\label{fig:f2}}
  \caption{AGNet predictions for SMBH redshift (left) and AGNet predictions for SMBH mass (right).  Ground truth shown by 1:1 black line with network predictions as the scatter. Trained over $\sim$7000 quasars using features from original light time series as well as spectroscopic redshift and $M_i$ from SDSS Stripe 82 (\S \ref{subsubsec:s82}) for mass estimations.  Optimized using SmoothL1 loss and AdamW optimizer for 50 epochs of training.}
\end{figure}

\section{Results}\label{sec:result}

\subsection{Network Performance on Redshift}\label{subsubsec:z_results}

Following \cite{Pasquet-Itam2018}, we compare our network performance on redshift estimation. We use \emph{ugriz} bands and colors as features for predicting redshift following the architecture outlined above.  Our best performance gives a RMSE = $0.355$ and a $R^2 = 0.759$.

We notice considerable deviation in network prediction for low redshift ($z < 1$).  One possible explanation for this is degeneracy in the color of quasars at redshift $z < 1$ and quasars at redshift $1 < z < 2$ \cite{Richards_2001}.

\subsection{Network Performance on Mass}

Using features from $\sim$10,000 spectroscopically confirmed quasars from the Sloan Digital Sky Survey, we achieve a RMSE = $0.357$ and a $R^2 = 0.418$. For context, this is already comparable to the systematic uncertainty in the spectroscopic mass ``ground truth'' estimate \cite{Shen2013}.
We also test mass predictions using information from only the original quasar light time series (see Figure \ref{fig:flowchart}). To do this, we use our AGNet architecture to predict $M_i$ using $ugriz$ colors, as well as the redshift predictions (\S \ref{subsubsec:z_results}) as testing data in predicting mass. In this test, we achieve RMSE = $0.401$ and a $R^2 = 0.272$.

\subsection{Comparison to K-Nearest Neighbors}

We now compare AGNet performance to a K-Nearest Neighbors (KNN) algorithm following \cite{Pasquet-Itam2018}. We follow the same features and preprocessing as our AGNet implementation, with an ideal K value of $K=19$. For quasar redshift, we achieve a RMSE = $0.386$ and for mass we achieve a RMSE = $0.370$. Using AGNet predicted redshift (z*) and $M_i$ values, KNN achieves RMSE = $0.398$, which is comparable to AGNet performance. The KNN performance suggests that spectral and time series features of limited data cannot be extracted as efficiently by traditional ML methods. 

\begin{table}[h]
\begin{tabular}{|p{3cm}||p{3cm}|p{3cm}||p{1.25cm}|p{1.25cm}|}
 \hline
 \multicolumn{5}{|c|}{Summary Statistics} \\
 \hline
 ML algorithm & Features & Parameters & RMSE & $R^2$\\
 \hline
 AGNet & colors and bands & redshift (z) & 0.355 & 0.759\\
 KNN   & colors and bands &  redshift (z) & 0.386 & 0.715\\
 \hline
 AGNet (w spec-z) & colors, $\tau$, $\sigma$, $M_i$, z & SMBH mass & 0.357 & 0.425\\
 KNN   (w spec-z) & colors, $\tau$, $\sigma$, $M_i$, z & SMBH mass & 0.370  & 0.369\\
 \hline
 AGNet (w/o spec-z) & colors, $\tau$, $\sigma$, $M_i$, z* & SMBH mass & 0.401 &  0.272\\
 KNN   (w/o spec-z) & colors, $\tau$, $\sigma$, $M_i$, z* & SMBH mass & 0.398 &  0.283\\
 \hline
\end{tabular}
\caption{Summary Statistics \label{tab:summary_stats}}
\end{table}

\section{Discussion and Future work}

We have shown that with photometric light curves, our AGNet pipeline provides a fast and automatic way to predict SMBH mass and redshift. The neural network is able to approximate a function from the given features to predict the desired parameters. It is able to learn from information provided from quasar light time series without expensive spectroscopic spectra.

To improve our model, we plan on implementing negative log likelihood loss to quantify uncertainties in our network predictions \cite{levasseur2017uncertainties}. We will also explore additional time series features for AGNet to learn from outside of $\sigma$ and $\tau$ using FATS \cite{nun2015fats}.
We will expand our work with larger data sets in future, such as the Dark Energy Survey Supernova Fields \cite{kessler2015difference} and the Vera Rubin Observatory. More robust and high quality data may improve our results, and may allow us to utilize more advanced CNN architectures and transfer learning techniques.  

\section*{Broader Impact}

We will open source our codes along with journal publication. Within the astronomy community, this work has potential for data analysis not only on AGN light curves, but also other transients detected by large-scale multi-band sky surveys (e.g. Vera Rubin Observatory). 

For broader impact on the public, we provide tools applicable for data analysis (parameter estimation) on time series data with large gaps (e.g. climate science, biology, medical diagnosis).

\begin{ack}

We thank the anonymous referees for helpful comments. This work utilizes resources supported by the National Science Foundation’s Major Research Instrumentation program, grant No.1725729, as well as the University of Illinois at Urbana-Champaign. We thank Dr. Volodymyr Kindratenko and Dr. Dawei Mu at the National Center for Supercomputing Applications (NCSA) for their assistance with the GPU cluster used in this work and helpful comments. JYYL, SP, and XL acknowledge support from the NCSA fellowship. SP and DP acknowledge support from the NCSA SPIN and NSF REU INCLUSION programs.

\end{ack}

\small

\bibliographystyle{unsrt}
\bibliography{mybib}

\begin{thebibliography}{10}

\bibitem{KormendyHo2013}
John Kormendy and Luis~C Ho.
\newblock Coevolution (or not) of supermassive black holes and host galaxies.
\newblock {\em Annual Review of Astronomy and Astrophysics}, 51:511--653, 2013.

\bibitem{Schmidt1968}
Maarten Schmidt.
\newblock Space distribution and luminosity functions of quasi-stellar radio
  sources.
\newblock {\em The Astrophysical Journal}, 151:393, 1968.

\bibitem{Rees1984}
Martin~J Rees.
\newblock Black hole models for active galactic nuclei.
\newblock {\em Annual review of astronomy and astrophysics}, 22(1):471--506,
  1984.

\bibitem{Soltan1982}
Andrzej Soltan.
\newblock Masses of quasars.
\newblock {\em Monthly Notices of the Royal Astronomical Society},
  200(1):115--122, 1982.

\bibitem{Banados2018}
Eduardo Ba{\~n}ados, Bram~P Venemans, Chiara Mazzucchelli, Emanuele~P Farina,
  Fabian Walter, Feige Wang, Roberto Decarli, Daniel Stern, Xiaohui Fan,
  Frederick~B Davies, et~al.
\newblock An 800-million-solar-mass black hole in a significantly neutral
  universe at a redshift of 7.5.
\newblock {\em Nature}, 553(7689):473--476, 2018.

\bibitem{King2014}
Anthea~L King, Tamara~M Davis, KD~Denney, Marianne Vestergaard, and D~Watson.
\newblock High-redshift standard candles: predicted cosmological constraints.
\newblock {\em Monthly Notices of the Royal Astronomical Society},
  441(4):3454--3476, 2014.

\bibitem{Dultzin2020}
Paola Marziani, Deborah Dultzin, Mauro D'Onofrio, JA~De~Diego~Onsurbe, Alenka
  Negrete, Ascensi{\'o}n Del~Olmo, Mary~Loli Mart{\'\i}nez-Aldama, Edi Bon,
  Natasa Bon, and Giovanna~Maria Stirpe.
\newblock Extreme quasars as distance indicators in cosmology.
\newblock {\em Frontiers in Astronomy and Space Sciences}, 6:80, 2019.

\bibitem{Shen2013}
Yue Shen.
\newblock The mass of quasars.
\newblock {\em arXiv preprint arXiv:1302.2643}, 2013.

\bibitem{rakshit2020spectral}
Suvendu Rakshit, CS~Stalin, and Jari Kotilainen.
\newblock Spectral properties of quasars from sloan digital sky survey data
  release 14: The catalog.
\newblock {\em The Astrophysical Journal Supplement Series}, 249(1):17, 2020.

\bibitem{Ivezic2019}
{\v{Z}}eljko Ivezi{\'c}, Steven~M Kahn, J~Anthony Tyson, Bob Abel, Emily
  Acosta, Robyn Allsman, David Alonso, Yusra AlSayyad, Scott~F Anderson, John
  Andrew, et~al.
\newblock Lsst: from science drivers to reference design and anticipated data
  products.
\newblock {\em The Astrophysical Journal}, 873(2):111, 2019.

\bibitem{Cabrera-Vives2017}
Guillermo Cabrera-Vives, Ignacio Reyes, Francisco F{\"o}rster, Pablo~A
  Est{\'e}vez, and Juan-Carlos Maureira.
\newblock Deep-hits: Rotation invariant convolutional neural network for
  transient detection.
\newblock {\em arXiv preprint arXiv:1701.00458}, 2017.

\bibitem{Charnock2017}
Tom Charnock and Adam Moss.
\newblock Deep recurrent neural networks for supernovae classification.
\newblock {\em The Astrophysical Journal Letters}, 837(2):L28, 2017.

\bibitem{Kim2017}
Edward~J Kim and Robert~J Brunner.
\newblock Star-galaxy classification using deep convolutional neural networks.
\newblock {\em Monthly Notices of the Royal Astronomical Society}, page
  stw2672, 2016.

\bibitem{George2018a}
Daniel George and EA~Huerta.
\newblock Deep learning for real-time gravitational wave detection and
  parameter estimation: Results with advanced ligo data.
\newblock {\em Physics Letters B}, 778:64--70, 2018.

\bibitem{Huang2018}
Xin Huang, Huaning Wang, Long Xu, Jinfu Liu, Rong Li, and Xinghua Dai.
\newblock Deep learning based solar flare forecasting model. i. results for
  line-of-sight magnetograms.
\newblock {\em The Astrophysical Journal}, 856(1):7, 2018.

\bibitem{Lanusse2018}
Fran{\c{c}}ois Lanusse, Quanbin Ma, Nan Li, Thomas~E Collett, Chun-Liang Li,
  Siamak Ravanbakhsh, Rachel Mandelbaum, and Barnab{\'a}s P{\'o}czos.
\newblock Cmu deeplens: deep learning for automatic image-based galaxy--galaxy
  strong lens finding.
\newblock {\em Monthly Notices of the Royal Astronomical Society},
  473(3):3895--3906, 2018.

\bibitem{Ribli2018}
Dezs{\H{o}} Ribli, B{\'a}lint~{\'A}rmin Pataki, and Istv{\'a}n Csabai.
\newblock An improved cosmological parameter inference scheme motivated by deep
  learning.
\newblock {\em Nature Astronomy}, 3(1):93--98, 2019.

\bibitem{Burke2019}
Colin~J Burke, Patrick~D Aleo, Yu-Ching Chen, Xin Liu, John~R Peterson, Glenn~H
  Sembroski, and Joshua Yao-Yu Lin.
\newblock Deblending and classifying astronomical sources with mask r-cnn deep
  learning.
\newblock {\em Monthly Notices of the Royal Astronomical Society},
  490(3):3952--3965, 2019.

\bibitem{yao2020feature}
Joshua Yao-Yu~Lin, George~N Wong, Ben~S Prather, and Charles~F Gammie.
\newblock Feature extraction on synthetic black hole images.
\newblock {\em arXiv e-prints}, pages arXiv--2007, 2020.

\bibitem{Pasquet-Itam2018}
Johanna Pasquet-Itam and J{\'e}r{\^o}me Pasquet.
\newblock Deep learning approach for classifying, detecting and predicting
  photometric redshifts of quasars in the sloan digital sky survey stripe 82.
\newblock {\em Astronomy \& Astrophysics}, 611:A97, 2018.

\bibitem{york2000sloan}
Donald~G York, J~Adelman, John~E Anderson~Jr, Scott~F Anderson, James Annis,
  Neta~A Bahcall, JA~Bakken, Robert Barkhouser, Steven Bastian, Eileen Berman,
  et~al.
\newblock The sloan digital sky survey: Technical summary.
\newblock {\em The Astronomical Journal}, 120(3):1579, 2000.

\bibitem{tachibana2020deep}
Yutaro Tachibana, Matthew~J Graham, Nobuyuki Kawai, SG~Djorgovski, Andrew~J
  Drake, Ashish~A Mahabal, and Daniel Stern.
\newblock Deep modeling of quasar variability.
\newblock {\em arXiv preprint arXiv:2003.01241}, 2020.

\bibitem{kelly2009variations}
Brandon~C Kelly, Jill Bechtold, and Aneta Siemiginowska.
\newblock Are the variations in quasar optical flux driven by thermal
  fluctuations?
\newblock {\em The Astrophysical Journal}, 698(1):895, 2009.

\bibitem{macleod2010modeling}
Ch~L MacLeod, {\v{Z}}~Ivezi{\'c}, CS~Kochanek, S~Koz{\l}owski, B~Kelly,
  E~Bullock, A~Kimball, B~Sesar, D~Westman, K~Brooks, et~al.
\newblock Modeling the time variability of sdss stripe 82 quasars as a damped
  random walk.
\newblock {\em The Astrophysical Journal}, 721(2):1014, 2010.

\bibitem{nun2015fats}
Isadora Nun, Pavlos Protopapas, Brandon Sim, Ming Zhu, Rahul Dave, Nicolas
  Castro, and Karim Pichara.
\newblock Fats: Feature analysis for time series.
\newblock {\em arXiv preprint arXiv:1506.00010}, 2015.

\bibitem{Shen2011}
Yue Shen, Gordon~T Richards, Michael~A Strauss, Patrick~B Hall, Donald~P
  Schneider, Stephanie Snedden, Dmitry Bizyaev, Howard Brewington, Viktor
  Malanushenko, Elena Malanushenko, et~al.
\newblock A catalog of quasar properties from sloan digital sky survey data
  release 7.
\newblock {\em The Astrophysical Journal Supplement Series}, 194(2):45, 2011.

\bibitem{agarap2018relu}
Abien Fred~M. Agarap.
\newblock Deep learning using rectified linear units (relu).
\newblock {\em arXiv:1803.08375}, 2018.

\bibitem{kingma2014adam}
Diederik~P Kingma and Jimmy Ba.
\newblock Adam: A method for stochastic optimization.
\newblock {\em arXiv preprint arXiv:1412.6980}, 2014.

\bibitem{loshchilov2017decoupled}
Ilya Loshchilov and Frank Hutter.
\newblock Decoupled weight decay regularization.
\newblock {\em arXiv preprint arXiv:1711.05101}, 2017.

\bibitem{he2016deep}
Kaiming He, Xiangyu Zhang, Shaoqing Ren, and Jian Sun.
\newblock Deep residual learning for image recognition.
\newblock In {\em Proceedings of the IEEE conference on computer vision and
  pattern recognition}, pages 770--778, 2016.

\bibitem{tan2019efficientnet}
Mingxing Tan and Quoc~V Le.
\newblock Efficientnet: Rethinking model scaling for convolutional neural
  networks.
\newblock {\em arXiv preprint arXiv:1905.11946}, 2019.

\bibitem{Richards_2001}
Gordon~T. Richards, Michael~A. Weinstein, Donald~P. Schneider, Xiaohui Fan,
  Michael~A. Strauss, Daniel~E. Vanden~Berk, James Annis, Scott Burles,
  Emily~M. Laubacher, Donald~G. York, and et~al.
\newblock Photometric redshifts of quasars.
\newblock {\em The Astronomical Journal}, 122(3):1151–1162, Sep 2001.

\bibitem{levasseur2017uncertainties}
Laurence~Perreault Levasseur, Yashar~D Hezaveh, and Risa~H Wechsler.
\newblock Uncertainties in parameters estimated with neural networks:
  Application to strong gravitational lensing.
\newblock {\em The Astrophysical Journal Letters}, 850(1):L7, 2017.

\bibitem{kessler2015difference}
R~Kessler, J~Marriner, M~Childress, R~Covarrubias, CB~D’Andrea, DA~Finley,
  J~Fischer, Ryan~Joseph Foley, D~Goldstein, RR~Gupta, et~al.
\newblock The difference imaging pipeline for the transient search in the dark
  energy survey.
\newblock {\em The Astronomical Journal}, 150(6):172, 2015.

\end{thebibliography}

\end{document}